\documentclass[12pt,a4paper,final]{iopart}

\usepackage{iopams}
\usepackage{graphicx}
\usepackage[breaklinks=true,colorlinks=true,linkcolor=blue,urlcolor=blue,citecolor=blue]{hyperref}

\newcommand{\arcsh}{\textrm{arcsinh}}  %
\newcommand{\arcth}{\textrm{arctanh}}  %

\eqnobysec

\begin{document}

\title[$X_m$ EOP, PDEM Scarf I potential and its coherent states revivals]{Revival structures of coherent states for $X_m$ exceptional
orthogonal polynomials of the Scarf I potential within position-dependent effective mass}

\author{Sid-Ahmed Yahiaoui and Mustapha Bentaiba\footnote{Author to whom any correspondence should be addressed.}}
\address{LPTHIRM, D\'epartement de physique, Facult\'e des sciences, Universit\'e Sa\^ad DAHLAB-Blida~1, B.P.~270 Route de Soum\^aa, 09\,000 Blida, Algeria}
\eads{\mailto{s$\_$yahiaoui@univ-blida.dz},\,\,\mailto{bentaiba@univ-blida.dz}}

\begin{abstract}
The revival structures for the $X_m$ exceptional orthogonal polynomials of the Scarf I potential endowed with position-dependent effective mass is studied in the context of the generalized Gazeau-Klauder coherent states. It is shown that in the case of the constant mass, the
deduced coherent states mimic full and fractional revivals phenomena. However in the case of position-dependent effective mass, although
full revivals take place during their time evolution, there is no fractional revivals as defined in the common sense. These properties are illustrated numerically by means of some specific profile mass functions, with and without singularities. We have also observed a close connection between the coherence time $\tau^{(m)}_{\rm coh}$ and the mass parameter $\lambda$.
\end{abstract}

\vspace{2pc}

\section{Introduction}%

\noindent It is well known that when working with quantum systems subjected to interact with a given interaction, usually considerations
require to identify the mass-term with the concept of effective mass. In a way, such quantum system becomes position-dependent effective mass (PDEM). In recent years, the study of quantum system endowed with PDEM has become one of the active subjects of research due to its
relevance in describing the properties of a wide variety of physical problems, such as quantum wells, wires and dots \cite{1}, and
semiconductor heterostructures \cite{2}. We can also find their applications in many others fields, such as the effective interactions in nuclear physics \cite{3}, curved spaces \cite{4}, $\mathcal PT$-symmetry \cite{5,6}, coherent states \cite{7,8,9,10,11}, and in the context of the Wigner's distribution functions \cite{12,13,14}.\\
\indent Such systems stimulated a lot of work in mathematical physics for finding the exact solutions for the PDEM Schr\"odinger equation
(PDEM SE). This quest has been addressed by many approaches and from different point of view; we may quote for instance the factorization method \cite{15}, supersymmetric quantum mechanics (SUSY QM) and the related shape-invariant potentials \cite{16}, and Lie algebra \cite{17,18,19}. The point canonical transformation (PCT) is one of these methods \cite{20} which consists to convert SE into a second-order differential equation, whose solutions are often expressed in terms of the classical orthogonal polynomials (COP) \cite{21}, known to play a very important role in the construction of the bound-states in quantum mechanics.\\
\indent Apart from COP, the introduction of $X_m$ exceptional orthogonal polynomials ($X_m$ EOP) by G\'omez-Ullate {\it et al.}
\cite{22,23} and Quesne \cite{24} has been considered as a big advance in the understanding of mathematics and physics that can be brought about by such eigenfunctions. The $X_m$ EOP are the solutions of the second-order Sturm-Liouville eigenvalues problem with {\it rational
coefficients}, obtained from the eigenfunctions of exactly solvable systems which have a degree $m$ eigen-polynomial deformation. Thus they form complete and orthogonal polynomial sets generalizing COP of Hermite, Laguerre and Jacobi. The term {\it exceptional} is used to indicate that these polynomials start at degree $m$ ($m\geq1$) called {\it codimension}, instead of the degree 0 constant term, thus avoiding restrictions of Bochner's theorem. Recently, $X_m$ EOP have been studied in a lot of works \cite{25,26,27,28,29,30,31,32,33,34,35,36,37,38}, including PDEM systems \cite{39}.\\
\indent On the other hand quantum revivals \cite{40,41}, which are the fundamental realization of the time-dependent interference phenomena for bound-states with quantized energy spectra, arise when the wave-packet spreads inside the potential and reconstruct itself during a certain time $T_{\rm rev}$, called the revival time. The same also occurs at some integer multiples of $T_{\rm rev}$, i.e. $(p/q)T_{\rm rev}$, when the evolving wave-packet will break up into a set of mini-packets of its original form. However, aside from the revival dynamics studied for systems with constant mass which have been well understood in many works \cite{42,43,44,45,46,47,48,49,50}, comparatively we are aware of few papers that dealt with revival dynamics for PDEM \cite{51,52} which did not receive much attention.\\
\indent It is the objective of this paper to fill this gap and to study the revival dynamics of the generalized Gazeau-Klauder coherent states (GK CS) \cite{53} for $X_m$ EOP of the Scarf I potential in the specific case of PDEM which, as far as we know, have not been considered yet. So the quest for studying CS wave-packets confined in an effective potential, with a predetermined energy spectrum, is very interesting in hopes to see how the mass function $M(x)$ can affect their temporal evolution. Our analysis reveals that, although that full revival still takes place during their time evolution $T_{\rm rev}$, there is {\it no trace of fractional revivals} in opposition to the usual case of the constant mass (CM). We observe that not only full revivals are different but also depend closely on the profile of the mass function used. These results are illustrated numerically by means of two profiles of the mass functions, with and without singularities, and agree with those obtained in \cite{51}. We have also established the correspondence between the coherence time $\tau^{(m)}_{\rm coh}$ and the mass parameter $\lambda$, defined in the sense that is emphasized in \cite{54}.\\
\indent The organization of our paper is as follows. In section 2 we generate the PDEM version of exactly solvable potential already
obtained in \cite{55}, in the case of constant mass and associated with the Scarf I potential, for which their wavefunctions involve the exceptional $X_m$ Jacobi polynomials via the PCT approach. In section 3 we show that the obtained potentials are shape invariant in the back-ground of supersymmetric quantum mechanics. In section 4 we study the revival dynamics of Gazeau-Klauder CS wave-packets for the exceptional $X_m$ Scarf I potential in the cases where the mass is constant and position-dependent. The last section is devoted to our conclusion.

\section{Generation of new PDEM potentials via PCT and their exceptional $X_m$ Jacobi polynomials}%

\begin{figure}[h]
 \centering
 \includegraphics[width=15.7cm,height=7cm]{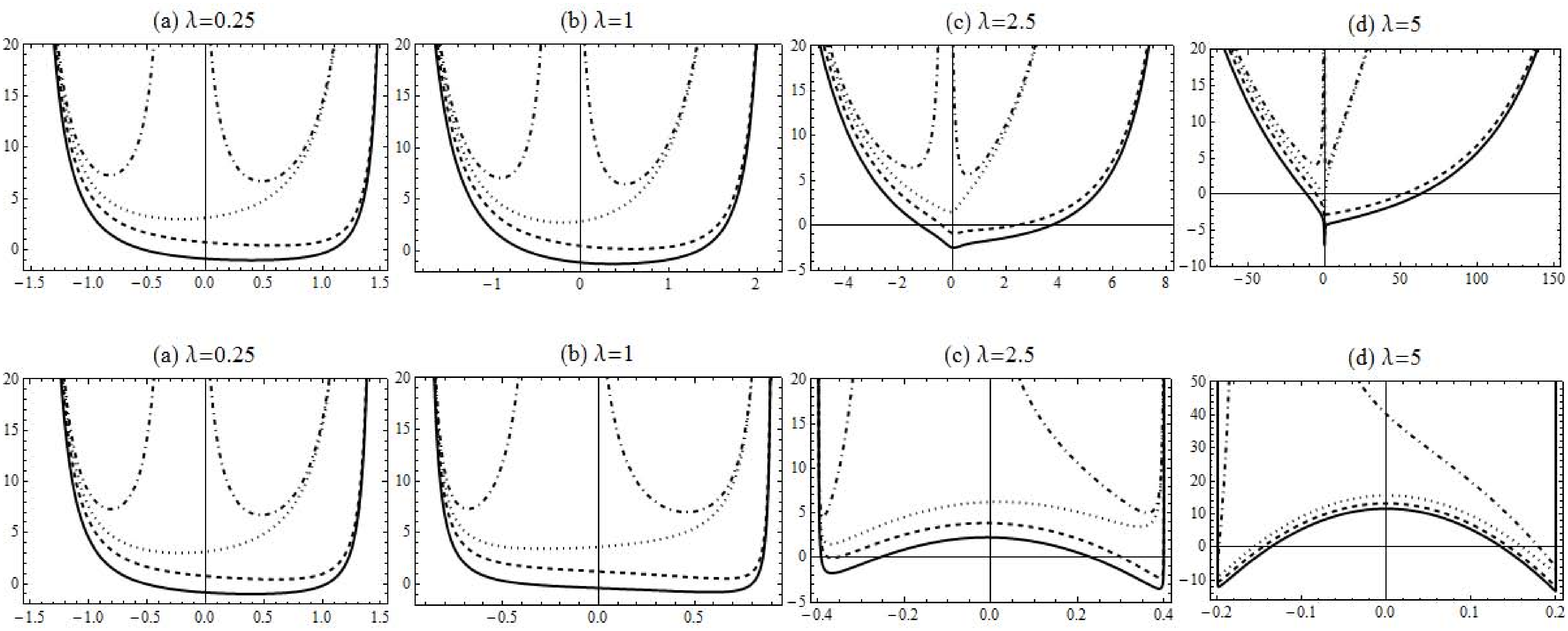}
 \caption{Top (resp. bottom). Plot of the effective potential \eref{2.15} plotted for $M_{\rm wos}(x)$ (resp. $M_{\rm ws}(x)$) for $\alpha=1$ and $\beta=2$: $m=0$ (solid line), $m=1$ (dashed line), $m=2$ (dotted line), and $m=3$ (dot-dashed line).} \label{figure1}
\end{figure}

\noindent Taking the natural units ($\hbar=m_0=1$) and using the ordering prescription adopted by BenDaniel and Duke \cite{56}, the
one-dimensional PDEM SE can be expressed as
\begin{eqnarray}\label{2.1}
  \left(-\frac{1}{2}\frac{\rmd^2}{\rmd x^2}+\frac{M'(x)}{2M(x)}\frac{\rmd}{\rmd x}+M(x)V(x)\right)\psi(x)=M(x)E_n\psi(x).
\end{eqnarray}
\indent Then by applying the following PCT, $\psi(x)=f(x)F\left(g(x)\right)$, to the eigenfunctions, it is not difficult to verify that
\eref{2.1} satisfies the second order differential equation \cite{20}
\begin{eqnarray}\label{2.2}
  \frac{\rmd^2F(g)}{\rmd g^2}+Q(g)\frac{\rmd F(g)}{\rmd g}+R(g)F(g)=0,
\end{eqnarray}
where $F(g)$ is some special function on $g(x)$. The functions $Q(g)$ and $R(g)$ are given by
\begin{eqnarray}
  Q(g) &=& \frac{g''(x)}{g'^2(x)}+\frac{2f'(x)}{f(x)g'(x)}-\frac{M'(x)}{M(x)g'(x)},\label{2.3} \\
  R(g) &=& \frac{f''(x)}{f(x)g'^2(x)}-\frac{M'(x)f'(x)}{M(x)f(x)g'^2(x)}+\frac{2M(x)}{g'^2(x)}\,\left(E_n-V(x)\right).\label{2.4}
\end{eqnarray}
\indent Integrating \eref{2.3}, we arrive to express $f(x)$ as
\begin{eqnarray}\label{2.5}
  f(x) &=& \sqrt{\frac{M(x)}{g'(x)}}\,\exp\left\{\frac{1}{2}\int^{g(x)}Q(g)\,\rmd g\right\},
\end{eqnarray}
and by inserting \eref{2.5} into \eref{2.4}, one can see that we obtain a system where the associated effective potential depends on the mass function
\begin{eqnarray}\label{2.6}
\fl  E_n-V_{\rm eff}(x) &=& \frac{g'^2(x)}{2M(x)}\,\left(R(g)-\frac{1}{2}\frac{\rmd Q(g)}{\rmd g}-\frac{1}{4}Q^2(g)\right)+\frac{1}{4M(x)}
\left(S(g')-S(M)\right),
\end{eqnarray}
where $S(z)=z''/z-3/2\,(z'/z)^2$ is the Schwartz derivative of the function $z(x)$ and the prime denotes the derivative with respect to $x$. It follows that the PDEM SE can be solved if the forms of $Q$ and $R$ are given for a mass function $M(x)$. In order to obtain the effective potential in the above equation, we impose that there must be a constant on the right-hand side of \eref{2.6} representing the bound-state energy spectrum $E_n$ on the left-hand side.\\
\indent From \eref{2.5} the solution of the eigenfunctions $\psi_n(x)$ are given by
\begin{eqnarray}\label{2.7}
  \psi_n(x)&\sim &\sqrt{\frac{M(x)}{g'(x)}}\,\exp\left\{\frac{1}{2}\int^{g(x)}Q(g)\,\rmd g\right\}\,F_n(g(x)),
\end{eqnarray}
up to a normalization constant. It is worth to note that all expressions reduce to the well known ones if the mass is taken to be constant, i.e., $M(x)=1$.\\
\indent In the remainder of the paper, we choose to work under the special function $F_n^{(m)}(g)$ to be the PDEM $X_m$ Jacobi polynomials $\widehat P_n^{(\alpha,\beta,m)}(g)$ studied in more details in \cite{29,55}, where $n\geq m$. This new family of orthogonal polynomials is orthonormal with respect to the weight function \cite{55}
\[
\widehat W^{(m)}(g)=\frac{(1-g)^\alpha(1+g)^\beta}{P_{m}^{(-\alpha-1,\beta-1)}(g)},\qquad \left(g\equiv g(x)\right),
\]
where $\widehat W^{(0)}(g)=W(g)$ is the weight function for the classical Jacobi polynomials and $-1\leq g(x)\leq+1$ in order that
$\mathbf {L}^2\left(g,\widehat W^{(m)}(g)\rmd g\right)$-orthonormality holds.\\
\indent Moreover these polynomials are related to the classical Jacobi orthogonal polynomials $P_{n}^{(\alpha,\beta)}(g)$ by the following relations
\begin{eqnarray}
\fl\widehat P_n^{(\alpha,\beta,0)}(g)=&P_n^{(\alpha,\beta)}(g), \label{2.8}\\
\fl\widehat P_n^{(\alpha,\beta,m)}(g)=&(-1)^m\Bigg[\frac{\alpha+\beta+j+1}{2(\alpha+j+1)}(g-1)P_m^{(-\alpha-1,\beta-1)}(g)
P_{j-1}^{(\alpha+2,\beta)}(g)\nonumber \\
&\qquad\quad+\frac{\alpha-m+1}{\alpha+j+1}P_m^{(-\alpha-2,\beta)}(g)P_{j}^{(\alpha+1,\beta-1)}(g)\Bigg], \quad (j=n-m\geq0) \label{2.9}
\end{eqnarray}
if and only if the following restrictions hold simultaneously \cite{22,55}
\numparts
\begin{eqnarray}
{\rm{\bf(R1)}}\quad && \beta\neq0,\alpha,\quad{\rm and}\quad\alpha-\beta-m+1\not\in\{0,1,2,\cdots,m-1\},   \label{2.10a}\\
{\rm{\bf(R2)}}\quad && \alpha>m-2,\quad{\rm and}\quad{\rm sgn}(\alpha-\beta+1)={\rm sgn}(\beta),           \label{2.10b}
\end{eqnarray}
\endnumparts
where {\rm sgn}($\cdot$) is the signum function. Under these conditions, the scalar product of the exceptional $X_m$ Jacobi polynomials
leads to orthogonality relation,
\begin{eqnarray}\label{2.11}
\fl &\int_{-1}^{1}&\frac{(1-g)^\alpha(1+g)^\beta}{\left[P_{m}^{(-\alpha-1,\beta-1)}(g)\right]^2}\,
\widehat P_n^{(\alpha,\beta,m)}(g)\widehat P_l^{(\alpha,\beta,m)}(g)\,\rmd g  \nonumber \\
\fl &&= \frac{2^{2s}(n-2m+\alpha+1)\Gamma(n+\beta+1)\Gamma(n-m+\alpha+2)}{(2n-2m+2s)(n-m+\alpha+1)^2
\Gamma(n-m+1)\Gamma(n-m+2s)}\,\delta_{n,l},
\end{eqnarray}
where $n,l\geq m$, $2s=\alpha+\beta+1$ and $\delta_{n,l}$ is the Kr\"onecker's symbol.\\
\indent Now for a fixed integer-parameter $m\geq1$ and real $\alpha,\beta>-1$, the functions $Q^{(m)}(g)$ and $R^{(m)}(g)$ can be generalized to the PDEM case and expressed in terms of classical Jacobi orthogonal polynomials $P_n^{(\alpha,\beta)}(g)$ through
\begin{eqnarray}
\fl Q^{(m)}(g)=(\alpha-\beta-m+1)\frac{P_{m-1}^{(-\alpha,\beta)}(g)}{P_{m}^{(-\alpha-1,\beta-1)}(g)}-
\frac{\alpha-\beta+(\alpha+\beta+2)g}{1-g^2},\label{2.12}\\
\fl R^{(m)}(g)=\frac{\beta(\alpha-\beta-m+1)}{1+g}\frac{P_{m-1}^{(-\alpha,\beta)}(g)}{P_{m}^{(-\alpha-1,\beta-1)}(g)}
+\frac{n^2+n(\alpha+\beta-2m+1)-2\beta m}{1-g^2},\label{2.13}
\end{eqnarray}
and substituting \eref{2.12} and \eref{2.13} into \eref{2.6}, we get after lengthy but straightforward computation
\begin{eqnarray}\label{2.14}
\fl  E_n^{(m)}-V^{(m)}_{\rm eff}(x)   &=& \frac{g'^2(x)}{4M(x)}\frac{2n(n-2m+\alpha+\beta+1)+2m(\alpha-3\beta-m+1)+\alpha+\beta+2}{1-g^2(x)} \nonumber \\
\fl &-& \frac{g'^2(x)}{8M(x)}\frac{\left(\alpha-\beta+(\alpha+\beta+2)g(x)\right)\left(\alpha-\beta+(\alpha+\beta-2)g(x)\right)}
{\left(1-g^2(x)\right)^2} \nonumber \\
\fl &+& \frac{g'^2(x)}{2M(x)}\frac{(\alpha-\beta-m+1)(\alpha+\beta+(\alpha-\beta+1)g(x))}{1-g^2(x)}
\frac{P_{m-1}^{(-\alpha,\beta)}(g(x))}{P_{m}^{(-\alpha-1,\beta-1)}(g(x))} \nonumber \\
\fl &-& \frac{g'^2(x)}{4M(x)}(\alpha-\beta-m+1)^2\left[\frac{P_{m-1}^{(-\alpha,\beta)}(g(x))}{P_{m}^{(-\alpha-1,\beta-1)}(g(x))}\right]^2 \nonumber \\
\fl &+& \frac{1}{4M(x)}\left[\frac{g'''(x)}{g'(x)}-\frac{3}{2}\frac{g''^2(x)}{g'^2(x)}\right]
       -\frac{1}{4M(x)}\left[\frac{M''(x)}{M(x)}-\frac{3}{2}\frac{M'^2(x)}{M^2(x)}\right].
\end{eqnarray}
\indent Equation \eref{2.14} can be solved by choosing an appropriate $g(x)$ in order to make the right-hand side having a constant dependent on $n$, and considering that the effective potential should be independent of $n$. Taking $g'^2(x)/(1-g^2(x))=cM(x)$, where $c>0$ is a constant, the solution of the above-mentioned differential equation $g(x)=\sin k\mu(x)\, (k=\sqrt c)$ leads to the construction of {\it infinite families of new PDEM Hermitian $X_m$ Scarf I potential} whose effective potentials $V^{(m)}_{\rm eff}(x)$, energy eigenvalues $E_n^{(m)}$ and eigenfunctions $\psi_n^{(m)}(x)$ are given by
\begin{eqnarray}
\fl V^{(m)}_{\rm eff}(x) =& \frac{k^2}{8}\left(2\alpha^2+2\beta^2-1\right)\sec^2\vartheta(x)-
\frac{k^2}{4}\left(\beta^2-\alpha^2\right)\sec \vartheta(x)\tan \vartheta(x) \nonumber \\
\fl &-\frac{k^2}{2}\left(\alpha-\beta-m+1\right)\left(\alpha+\beta
+(\alpha-\beta+1)\sin \vartheta(x)\right)\frac{P_{m-1}^{(-\alpha,\beta)}(\sin \vartheta(x))}{P_{m}^{(-\alpha-1,\beta-1)}(\sin \vartheta(x))} \nonumber \\
\fl &+\frac{k^2}{4}\left(\alpha-\beta-m+1\right)^2\cos^2 \vartheta(x)
\left[\frac{P_{m-1}^{(-\alpha,\beta)}(\sin \vartheta(x))}{P_{m}^{(-\alpha-1,\beta-1)}(\sin \vartheta(x))}\right]^2 \nonumber \\
\fl &+\frac{1}{4}\frac{\mu'''(x)}{\mu'^3(x)}-\frac{5}{8}\frac{\mu''^2(x)}{\mu'^4(x)}
-\frac{k^2}{8}\left[(\alpha+\beta+1)^2+4m(\alpha-3\beta-m+1)\right], \label{2.15} \\
\fl E_n^{(m)} \equiv&  k^2 e_n^{(m)} = \frac{k^2}{2}\,n(n-2m+\alpha+\beta+1), \label{2.16} \\
\fl \psi_n^{(m)}(x) =& N_n^{(m)} M^{1/4}(x)\,\frac{(1-\sin \vartheta(x))^{\frac{\alpha}{2}
+\frac{1}{4}}(1+\sin\vartheta(x))^{\frac{\beta}{2}+\frac{1}{4}}}{P_{m}^{(-\alpha-1,\beta-1)}(\sin \vartheta(x))}\,\widehat P_n^{(\alpha,\beta,m)}
(\sin\vartheta(x)), \label{2.17}
\end{eqnarray}
where $\vartheta(x)=k\mu(x)$, $N_n^{(m)}$ is the normalization constant and where we introduce the auxiliary mass function $\mu'(x)=\sqrt{M(x)}$. We expressly chose to deduce the effective potential in its shape given by \eref{2.15} and its associated energy spectra \eref{2.16}, so that the ground-state energy $E_0^{(m)}$ is chosen to be zero, in order to construct their associated coherent states {\it \`a la Gazeau-Klauder} in the next section.\\
\indent It is worth noting that the bound-state wavefunctions \eref{2.17} are physically acceptable if and only if the square-integrability condition fulfills the restriction $|\psi_n^{(m)}(x)|^2/\sqrt{M(x)}\rightarrow0$ at the end points $x_1$ and $x_2$ of the interval of the effective potential \eref{2.15}.Then in order to normalize \eref{2.17}, as in the case of usual CM, we must require that the auxiliary mass function $\mu(x)$ can be restricted to the region
\begin{eqnarray}\label{2.18}
  -1\leq g(x)\leq +1  \qquad\Rightarrow\qquad  -\frac{\pi}{2k}\leq\mu(x)\leq\frac{\pi}{2k},
\end{eqnarray}
which will be used later once we choose the profile of the mass function $M(x)$. Thus with the help of \eref{2.11}, the normalized constant $N_n^{(m)}$ is given by
\[
\fl N_n^{(m)}=\frac{k}{2^{s-1/2}}\sqrt{\frac{(n-m+s)\,(n-m+\alpha+1)^2\,\Gamma(n-m+1)\,\Gamma(n-m+2s)}
{(n-2m+\alpha+1)\,\Gamma(n-m+\alpha+2)\,\Gamma(n+\beta+1)}},
\]
where $s=(\alpha+\beta+1)/2$ and $\alpha,\beta>-1$. The new PDEM potentials \eref{2.15} are infinite, since each $m\geq1$ gives rise to {\it new exactly solvable PDEM potentials} which are all singular in the interval \eref{2.18}, due to the properties of the Jacobi polynomials.\\ \indent For $m=0$, we recognize the well-known PDEM Scarf I potential associated to the classical Jacobi polynomials \cite{20} (see for instance, equation (23b) therein), namely
\begin{eqnarray}\label{2.19}
 V^{(0)}_{\rm eff}(x) =& \frac{k^2}{8}\left(2\alpha^2+2\beta^2-1\right)\sec^2\vartheta(x)-\frac{k^2}{4}\left(\beta^2-\alpha^2\right)\sec\vartheta(x)\tan\vartheta(x) \nonumber \\
 &-\frac{k^2}{8}(\alpha+\beta+1)^2+\frac{1}{4}\frac{\mu'''(x)}{\mu'^3(x)}-\frac{5}{8}\frac{\mu''^2(x)}{\mu'^4(x)},
\end{eqnarray}
and all other potentials ($m\geq1$) are considered as extension of $V^{(0)}_{\rm eff}(x)$. Then it is obvious to interpret
$V^{(m)}_{\rm eff}(x)$ as the {\it PDEM rationally extended} Scarf I potential family.\\
\indent In \Fref{figure1} we have depicted the effective potential $V^{(m)}_{\rm eff}(x)$ given in \eref{2.15} plotted for two profile of the mass functions, i.e. with and without singularities \cite{59}
\begin{eqnarray}\label{2.20}
    M_{\rm wos}(x)=\frac{1}{1+(\lambda x)^2},\qquad{\rm and} \qquad M_{\rm ws}(x)=\frac{1}{\left(1-(\lambda x)^2\right)^2},
\end{eqnarray}
inside the interval \eref{2.18}, for even and odd $m$ up to 3 and for different values of the mass parameter $\lambda\in\mathbb R$.\\
\indent Due to the singularities of $M_{\rm ws}(x)$, we observe that gradually as $\lambda$ increases the shape of $V^{(m)}_{\rm eff}(x)$ tend to gather near the {\it classical turning points} $x_\pm=1/\lambda$ of the well. Contrary, the case $M_{\rm wos}(x)$ reveals that $V^{(m)}_{\rm eff}(x)$ are more extended along the whole real line $\mathbb R$ and become more sharper at the vicinity of $x_0=0$, as $\lambda$ increases.

\section{Supersymmetry and shape-invariant approach}%

\noindent It is well-known that supersymmetric quantum mechanics (SUSY QM) has been successfully applied to obtained exact solutions
of Schr\"odinger equation, which deals with pairs of Hamiltonians $\mathcal H_{1,{\rm eff}}^{(m)}$ and $\mathcal H_{2,{\rm eff}}^{(m)}$
that have the same energy spectra, but different eigenfunctions. It has been shown that such pairs of Hamiltonians can be obtained through the concept of shape invariance, which is the sufficient condition for {\it exact solvability} and satisfying \cite{16}
\begin{eqnarray}\label{3.1}
  \mathcal V_{2,{\rm eff}}^{(m)}(x|\{\mathbf{a}_1\}) &=& \mathcal V_{1,{\rm eff}}^{(m)}(x|\{\mathbf{a}_2\})
                                                         +R^{(m)}(\{\mathbf{a}_1\}),
\end{eqnarray}
where $\{\mathbf{a}_i\}$, $(i=1,2,\cdots)$, is a set of parameters, $\{\mathbf{a}_{i+1}\}=f(\{\mathbf{a}_i\})$ is an arbitrary function describing the change of parameters and $R^{(m)}\left(\{\mathbf{a}_i\}\right)$ is independent of $x$. If these conditions are fulfilled, then the energy spectra of $\mathcal V_{1,{\rm eff}}^{(m)}(x)$ can be obtained algebraically
\begin{eqnarray}\label{3.2}
  E_{1,n}^{(m)} &=& \sum_{i=1}^n R^{(m)}(\{\mathbf{a}_i\}).
\end{eqnarray}
\indent In the light of the last section, we introduce a pair of operators $\widehat{\mathcal Q}_m$ and $\widehat{\mathcal Q}_m^\dagger$ and the associated superpotential $\mathcal W_m(x)$ \cite{59,60} through
\numparts
\begin{eqnarray}
  \widehat{\mathcal Q}_m &=& \frac{1}{\sqrt2}\left(\frac{1}{M^{1/4}(x)}\frac{\rmd}{\rmd x}\frac{1}{M^{1/4}(x)}+\mathcal W_m(x)\right), \label{3.3a} \\
  \widehat{\mathcal Q}_m^\dagger &=& \frac{1}{\sqrt2}\left(-\frac{1}{M^{1/4}(x)}\frac{\rmd}{\rmd x}\frac{1}{M^{1/4}(x)}+\mathcal W_m(x)\right), \label{3.3b}
\end{eqnarray}
\endnumparts
where the superpotential $\mathcal W_m(x)$ is defined in terms of the ground-state wavefunctions
$\psi_m^{(m)}(x)$, for $n=m$, as
\begin{eqnarray}\label{3.4}
  \psi_m^{(m)}(x) &\sim& \frac{1}{\sqrt{M(x)}}\,\exp\left\{-\int^{\mu(x)}\mathcal W_m(\eta)\,\rmd\mu(\eta)\right\}.
\end{eqnarray}
\indent The operators defined in \eref{3.3a} and \eref{3.3b} give rise to two effective partner Hamiltonians, namely,
$\widehat{\mathcal H}_{1,{\rm eff}}^{(m)}\equiv\widehat{\mathcal Q}_m^\dagger\widehat{\mathcal Q}_m$ and
$\widehat{\mathcal H}_{2,{\rm eff}}^{(m)}\equiv\widehat{\mathcal Q}_m\widehat{\mathcal Q}_m^\dagger$, which are given by
\numparts
\begin{eqnarray}
  \widehat{\mathcal H}_{1,{\rm eff}}^{(m)} &=&\frac{1}{2}\left[-\left(\frac{1}{M^{1/4}(x)}\frac{\rmd}{\rmd x}\frac{1}{M^{1/4}(x)}\right)^2                   +\mathcal W_m^2(x)-\frac{\mathcal W_m'(x)}{\sqrt{M(x)}}\right], \label{3.5a} \\
  \widehat{\mathcal H}_{2,{\rm eff}}^{(m)} &=&\frac{1}{2}\left[-\left(\frac{1}{M^{1/4}(x)}\frac{\rmd}{\rmd x}\frac{1}{M^{1/4}(x)}\right)^2                   +\mathcal W_m^2(x)+\frac{\mathcal W_m'(x)}{\sqrt{M(x)}}\right], \label{3.5b}
\end{eqnarray}
\endnumparts
where the relationship connecting the two effective partner potentials is given by
\begin{eqnarray}\label{3.6}
  \mathcal V_{2,{\rm eff}}^{(m)}(x) &=& \mathcal V_{1,{\rm eff}}^{(m)}(x)+\frac{1}{\sqrt{M(x)}}\mathcal W_m'(x),
\end{eqnarray}
which share the same energy spectrum, except the zero energy state, i.e. $E_{1,n+1}^{(m)}=E_{2,n}^{(m)}$ and $E_{1,0}^{(m)}=0$, for
$n=0,1,2,\cdots$ and $m=1,2,3,\cdots$.\\
\indent Due to \eref{2.16}, we identify the effective partner potential given in \eref{2.15} with $\mathcal V_{1,{\rm eff}}^{(m)}(x)$ and
the ground-state wavefunction can be deduced straightforwardly from \eref{2.17}
\begin{eqnarray}\label{3.7}
\fl \psi_m^{(m)}(x)=& N_m^{(m)} M^{1/4}(x)\,\frac{(1-\sin\vartheta(x))^{\frac{\alpha}{2}+\frac{1}{4}}
(1+ \sin\vartheta(x))^{\frac{\beta}{2}+\frac{1}{4}}}{P_{m}^{(-\alpha-1,\beta-1)}(\sin\vartheta(x))}\,\widehat P_m^{(\alpha,\beta,m)}(\sin\vartheta(x)),
\end{eqnarray}
where using \eref{2.9}, the exceptional $X_m$ Jacobi polynomials in \eref{3.7} are reduced in terms of the classical Jacobi polynomials as
\begin{eqnarray}\label{3.8}
  \widehat P_m^{(\alpha,\beta,m)}(\sin \vartheta(x)) &=& (-1)^m\left(1-\frac{m}{\alpha+1}\right)P_m^{(-\alpha-2,\beta)}(\sin\vartheta(x)).
\end{eqnarray}
\indent Making use of \eref{3.4}, the superpotential $\mathcal W_m(x)$ is given through
\begin{eqnarray}\label{3.9}
\fl  \mathcal W_m(x) = \frac{1}{4}\frac{M'(x)}{M^{3/2}(x)}-\frac{1}{\sqrt{M(x)}}\frac{\rmd}{\rmd x}\ln\psi_m^{(m)}(x) \nonumber \\
\fl  \qquad = \frac{k}{2}(\alpha-\beta)\sec\vartheta(x)+\frac{k}{2}(\alpha+\beta+1)\sec\vartheta(x)\tan\vartheta(x) \nonumber \\
\fl  \qquad-\frac{k}{2}(\alpha-\beta-m+1)\cos\vartheta(x)\left(\frac{P_{m-1}^{(-\alpha,\beta)}(\sin\vartheta(x))}{P_m^{(-\alpha-1,\beta-1)}
(\sin\vartheta(x))}-\frac{P_{m-1}^{(-\alpha-1,\beta-1)}(\sin\vartheta(x))}{P_m^{(-\alpha-2,\beta)}(\sin\vartheta(x))}\right)
\end{eqnarray}
where $\vartheta(x)=k\mu(x)$. Inserting \eref{3.9} into \eref{3.5a} and \eref{3.5b} and after some lengthy and algebraic manipulations, we obtain the simplified expression of the effective partner potentials
\numparts
\begin{eqnarray}
\fl  \mathcal V_{1,{\rm eff}}^{(m)}(x|\alpha,\beta) = \frac{k^2}{8}\left(2\alpha^2+2\beta^2-1\right)\sec^2\vartheta(x)-\frac{k^2}{4}
\left(\beta^2-\alpha^2\right)\sec\vartheta(x)\tan\vartheta(x) \nonumber \\
\fl \qquad\qquad\quad-\frac{k^2}{2}\left(\alpha-\beta-m+1\right)\left(\alpha+\beta+(\alpha-\beta+1)\sin\vartheta(x)\right)\frac{P_{m-1}^{(-\alpha,\beta)}
(\sin\vartheta(x))}{P_{m}^{(-\alpha-1,\beta-1)}(\sin\vartheta(x))} \nonumber \\
\fl \qquad\qquad\quad+\frac{k^2}{4}\left(\alpha-\beta-m+1\right)^2\cos^2\vartheta(x)\left[\frac{P_{m-1}^{(-\alpha,\beta)}(\sin\vartheta(x))}
{P_{m}^{(-\alpha-1,\beta-1)}(\sin\vartheta(x))}\right]^2 \nonumber \\
\fl \qquad\qquad\quad-\frac{k^2}{8}\left[(\alpha+\beta+1)^2+4m(\alpha-3\beta-m+1)\right], \label{3.10a} \\
\fl  \mathcal V_{2,{\rm eff}}^{(m)}(x|\alpha,\beta)=\frac{k^2}{8}\left(2(\alpha+1)^2+2(\beta+1)^2-1\right)\sec^2\vartheta(x) \nonumber \\
\fl \qquad\qquad-\frac{k^2}{4}\left((\beta+1)^2-(\alpha+1)^2\right)\sec\vartheta(x)\tan\vartheta(x) \nonumber \\
\fl \qquad\qquad-\frac{k^2}{2}\left(\alpha-\beta-m+1\right)\left(\alpha+\beta+2+(\alpha-\beta+1)\sin\vartheta(x)\right)
\frac{P_{m-1}^{(-\alpha-1,\beta+1)}(\sin\vartheta(x))}{P_{m}^{(-\alpha-2,\beta)}(\sin\vartheta(x))} \nonumber \\
\fl \qquad\qquad+\frac{k^2}{4}\left(\alpha-\beta-m+1\right)^2\cos^2\vartheta(x)\left[\frac{P_{m-1}^{(-\alpha-1,\beta+1)}(\sin\vartheta(x))}
{P_{m}^{(-\alpha-2,\beta)}(\sin\vartheta(x))}\right]^2 \nonumber \\
\fl \qquad\qquad-\frac{k^2}{8}\left[(\alpha+\beta+1)^2+4m(\alpha-3\beta-m+1)\right]. \label{3.10b}
\end{eqnarray}
\endnumparts
\indent We observe that the effective potential \eref{3.10a} matches with \eref{2.15}, apart from the omission of the Schwartz's derivative of the mass function due to our particular choice of the operators $\widehat{\mathcal Q}_m$ and $\widehat{\mathcal Q}_m^\dagger$ in \eref{3.3a} and \eref{3.3b}, respectively. Using \eref{3.1}, it is easy for the reader to check that \eref{3.10a} and \eref{3.10b} are connected to each other through the {\it translational shape invariant symmetry}, namely
\begin{eqnarray}\label{3.11}
  \mathcal V_{2,{\rm eff}}^{(m)}(x|\alpha,\beta) &=& \mathcal V_{1,{\rm eff}}^{(m)}(x|\alpha+1,\beta+1)+R^{(m)}(\alpha,\beta) \nonumber \\
   &=& \mathcal V_{1,{\rm eff}}^{(m)}(x|\alpha+1,\beta+1) + \frac{k^2}{2}(\alpha+\beta-2m+2),
\end{eqnarray}
where here the set of parameters $\{\mathbf{a}_i\}$ are defined by: $\{\mathbf{a}_1\}=(\alpha,\beta)$, $\{\mathbf{a}_2\}=(\alpha+1,\beta+1)$, and thus $\{\mathbf{a}_n\}=(\alpha+n-1,\beta+n-1)$. In view of \eref{3.2}, the bound-state energy eigenvalues of the effective potential $\mathcal V_{1,{\rm eff}}^{(m)}(x)$ are then given by
\begin{eqnarray}\label{3.12}
  E_{1,n}^{(m)} &=& \sum_{i=1}^n R^{(m)}(\{\mathbf{a}_i\}) \nonumber \\
                &=& \sum_{i=1}^n \frac{k^2}{2}(\alpha+\beta-2m+2i) \nonumber \\
                &=& \frac{k^2}{2}\,n(n+\alpha+\beta-2m+1),
\end{eqnarray}
which are just the energy eigenvalues deduced in \eref{2.16}, with the fact that $E_{1,0}^{(m)}=0$ as it was expected.

\section{Revival dynamics of Gazeau-Klauder CS for the extended PDEM Scarf I potential}%

\noindent Let us now adapt the material developed above to construct the Gazeau-Klauder CS (GK CS) \cite{53} for the extended Hermitian PDEM Scarf I potential given in \eref{2.15}. Such coherent states are parameterized by two real parameters $J$ and $\gamma$, and are defined by
\begin{eqnarray}\label{4.1}
  |\xi^{(m)}(x;J,\gamma)\rangle &=& \frac{1}{\mathcal N_m(J)}\sum_{n=0}^\infty\frac{J^{n/2}
  \,\exp\left\{-\rmi e_n^{(m)}\gamma\right\}}{\sqrt{\rho_n^{(m)}}} |\psi_n^{(m)}(x)\rangle,
\end{eqnarray}
where $\gamma=\omega t$. Here $e_n^{(m)}$ are the dimensionless non-degenerate energy eigenvalues \eref{2.16}, satisfying $e_{n+1}^{(m)}>e_n^{(m)}>e_{n-1}^{(m)}>\cdots>e_0^{(m)}\stackrel{\mbox{\tiny{\eref{2.16}}}}{=}0,\,(m\geq1)$,
and the parameter $\rho_n^{(m)}$ denotes the moments of probability distribution defined by $\rho_n^{(m)}=\prod_{i=1}^n e_n^{(m)}$, with $\rho_0^{(m)}=1$. The last parameter appearing in \eref{4.1} is the normalization constant given by
\begin{eqnarray}\label{4.2}
  \mathcal N_m(J) &=& \left(\sum_{n=0}^\infty\frac{J^n}{\rho_n^{(m)}}\right)^{1/2},
\end{eqnarray}
which is deduced from the normalization condition $\langle\xi^{(m)}(x;J,\gamma)|\xi^{(m)}(x;J,\gamma)\rangle=1$, where
$0<J<R=\lim_{n\rightarrow +\infty}\sup\sqrt[n]{\rho_n^{(m)}}$ and $R$ denotes the radius of convergence. Under these considerations, the
moments $\rho_n^{(m)}$ and the squared of normalization constant $\mathcal N_m(J)$ are given by
\begin{eqnarray}
  \rho_n^{(m)}      = \frac{n!}{2^n}\frac{\Gamma(n+2\sigma+1)}{\Gamma(2\sigma+1)}, \label{4.3} \\
  \mathcal N^2_m(J) = (2J)^{-\sigma}\Gamma(2\sigma+1)I_{2\sigma}(2\sqrt{2J}), \label{4.4}
\end{eqnarray}
and applying the Stirling's approximation to \eref{4.3} we get $R=\infty$. Here $\mathbb R\ni\sigma=s-m$ and $I_{2\sigma}(\cdot)$ are the modified Bessel functions of the first kind \cite{57}.\\
\indent So that, the Gazeau-Klauder CS \eref{4.1} are reduced to
\begin{eqnarray}\label{4.5}
\fl  |\xi^{(m)}(x;J,t)\rangle &=& \frac{(2J)^{\sigma/2}}{\sqrt{I_{2\sigma}(2\sqrt{2J})}}
     \sum_{n=0}^\infty\frac{(2J)^{n/2}\exp\left\{-\rmi\,\omega\,\frac{n(n+2\sigma)}{2}\,t\right\}}{\sqrt{n!\,\Gamma(n+2\sigma+1)}}\,
     |\psi_n^{(m)}(x)\rangle,
\end{eqnarray}
where $\psi_n^{(m)}(x)$ are given by \eref{2.17}. It is well known that the concept of quantum revivals arises from the weighting
probabilities $|c_n^{(m)}|^2$ for the general wave-packet, i.e.,
\begin{eqnarray}\label{4.6}
  |\Psi_n^{(m)}(x,t)\rangle &=& \sum_{n=0}^\infty c_n^{(m)} |\psi_n^{(m)}(x)\rangle,
\end{eqnarray}
where $\sum_{n=0}^\infty |c_n^{(m)}|^2=1$. So, when $|\Psi_n^{(m)}(x,t)\rangle$ in \eref{4.6} play the role of our Gazeau-Klauder CS \eref{4.5}, then the weighting distribution depends on $J$ as
\begin{eqnarray}\label{4.7}
  |c_n^{(m)}(J)|^2 &\equiv& \frac{J^n}{\mathcal N^2_m(J)\rho_n^{(m)}} = \frac{(2J)^{n+\sigma}}{n!\,\Gamma(n+2\sigma+1)I_{2\sigma}(2\sqrt{2J})}.
\end{eqnarray}

\begin{figure}[h]
 \centering
 \includegraphics[width=16cm,height=3cm]{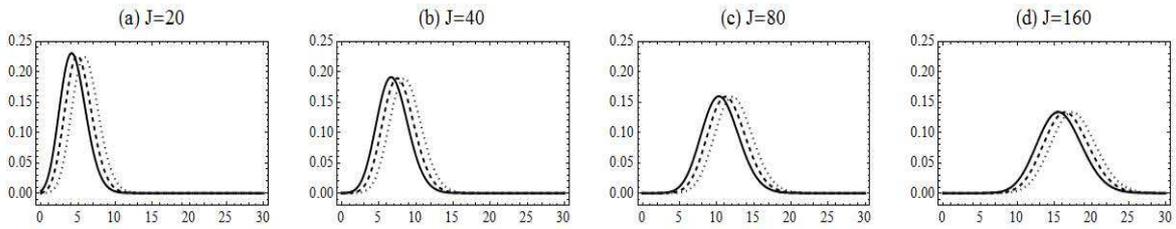}
 \caption{Plot of the weighting distribution $|c_n^{(m)}(J)|^2$ against $n$ for $\alpha=1$ and $\beta=2$, where $m=0$ (solid line), $m=1$ (dashed line), and $m=2$ (dotted line).} \label{figure2}
\end{figure}

\indent In \Fref{figure2} we display the curves of $|c_n^{(m)}(J)|^2$ as a function of quantum number $n$ for various values of $J$ and $m$. It is clear that all frames show a Gaussian-shaped function for the weighting distribution. We can observe that gradually as $J$
increases, the weighting distributions become more and more stretched and are less peaked with a slight shift to the right localized around a mean value $\overline{n}^{(m)}\simeq \langle n^{(m)} \rangle$.\\
\indent On the other hand, the mean and the variance values of the number operator $\widehat N_m$ are used to characterize the statistical features of the quantum system, which can be evaluated by means of the moments of probability. By making use of \eref{4.7}, a straightforward analytical calculation yields
\numparts
\begin{eqnarray}
  \langle n\rangle &\equiv& \sum_{n=0}^\infty n|c_n^{(m)}(J)|^2
  = \sqrt{2J}\,\frac{I_{2\sigma+1}(2\sqrt{2J})}{I_{2\sigma}(2\sqrt{2J})}, \label{4.8a} \\
  \langle n^2\rangle &\equiv& \sum_{n=0}^\infty n^2|c_n^{(m)}(J)|^2
  = 2J\,\frac{I_{2\sigma+2}(2\sqrt{2J})}{I_{2\sigma}(2\sqrt{2J})}+\sqrt{2J}\,\frac{I_{2\sigma+1}(2\sqrt{2J})}{I_{2\sigma}(2\sqrt{2J})},
  \label{4.8b}
\end{eqnarray}
\endnumparts
in order to display the Mandel parameter $Q_{\rm M}^{(m)}(J)$
\begin{eqnarray}\label{4.9}
  Q_{\rm M}^{(m)}(J) &\equiv& \frac{\langle n^2\rangle-\langle n\rangle^2}{\langle n\rangle}-1
                  = \sqrt{2J}\left(\frac{I_{2\sigma+2}(2\sqrt{2J})}{I_{2\sigma+1}(2\sqrt{2J})}
                  -\frac{I_{2\sigma+1}(2\sqrt{2J})}{I_{2\sigma}(2\sqrt{2J})}\right).
\end{eqnarray}
\indent The behavior of CS may be characterized through the Mandel parameter. It is an efficient way to characterize non-classical states
which have no classical analog. The case $Q_{\rm M}^{(m)}(J)=0$ coincides with the definition of CS, while for $Q_{\rm M}^{(m)}(J)<0$ and
$Q_{\rm M}^{(m)}(J)>0$ correspond to the sub-Poissonian and super-Poissonian statistics, respectively.

\begin{figure}[h]
 \centering
 \includegraphics[width=16cm,height=12cm]{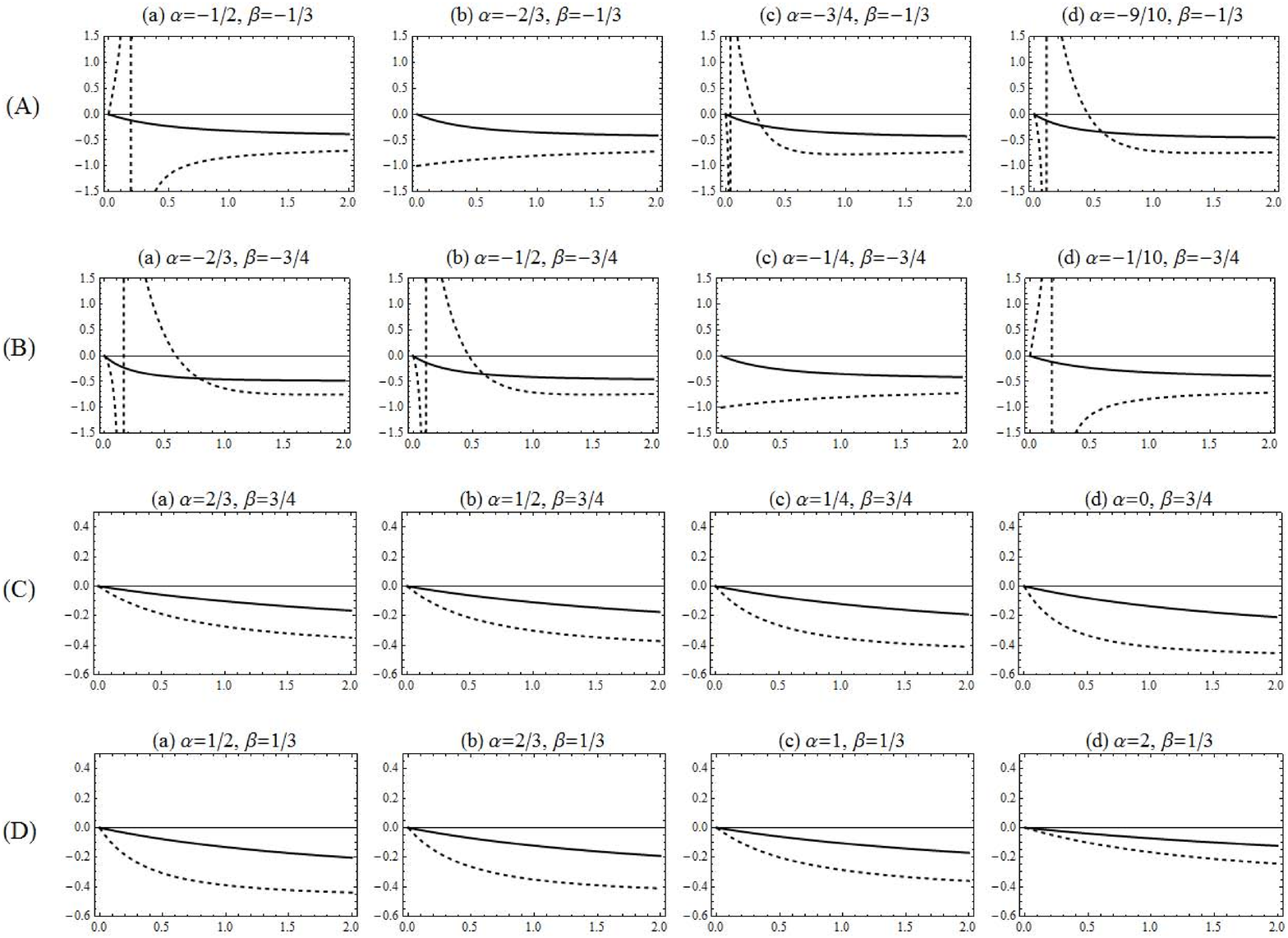}
 \caption{The Mandel parameter $Q_{\rm M}^{(m)}(J)$ given by \eref{4.9} against $J$ for fixed value for $\beta$ and varying $\alpha$ satisfying the conditions (4.10). The parameter $m$ refers to: $m=0$ (solid line) and $m=1$ (dashed line).} \label{figure3}
\end{figure}

\indent In \Fref{figure3} we display the behavior of the Mandel parameter \eref{4.9} against $J$ in terms of different values of the parameters $\alpha$, $\beta$ and $m=0,1$. Solving simultaneously \eref{2.10a} and \eref{2.10b} for $m=1$ give rise to four different cases depicted in \Fref{figure3} by letters (A), (B), (C), and (D), respectively, i.e.,
\numparts
\begin{eqnarray}
  (A):\quad -1<\beta<0, \quad 1+\alpha>0, \quad{\rm and}\quad \alpha<\beta \label{4.10a} \\
  (B):\quad -1<\beta<0, \quad \alpha\leq0, \qquad\,\,\,{\rm and}\quad \alpha>\beta \label{4.10b} \\
  (C):\quad \beta>0, \qquad\quad\,\,\: \alpha<\beta, \qquad\,\,{\rm and}\quad \alpha\geq0 \label{4.10c} \\
  (D):\quad \beta>0, \qquad\quad\,\,\: \alpha>\beta \label{4.10d}
\end{eqnarray}
\endnumparts
keeping in mind that $\alpha,\beta>-1$.\\
\indent It is clear that the case $m=0$ (solid lines), representing the classical Jacobi polynomials, exhibits the sub-Poissonian photon statistics, no matter what values are attributed to the parameters $\alpha$ and $\beta$. However the case $m=1$ (dashed lines), corresponding to the exceptional $X_1$ Jacobi polynomials, has a completely different behavior compared to that of the case $m=0$. At this stage, a few remarks are worth mentioning:
\begin{enumerate}
  \item For the fixed value $\beta=-\frac{1}{3}$, the state in A(a) starts with a super-Poissonian behavior for a short range in $J$ and becomes sub-Poissonian for $J\simeq0.2$. Gradually as $\alpha$ decreases the trend reverses, and {\it a kind of transition takes place} so that the states in A(c) (resp. A(d)) start at slightly sub-Poissonian, increase to super-Poissonian at $J\simeq0.05$ (resp. $J\simeq0.10$) and then decrease very fast to become sub-Poissonian at $J\simeq0.25$ (resp. $J\simeq0.45$).
  \item For $\beta=-\frac{3}{4}$, it is found that the states B(a) (resp. B(b)) acquire at the beginning a sub-Poissonian behavior for a short range of $J$, become super-Poissonian at $J\simeq0.15$ (resp. $J\simeq0.11$), and decrease very fast to sub-Poissonian state at $J\simeq0.6$ (resp. $J\simeq0.45$). In the frame B(d), the process is completely reversed.
  \item As seen in frames (C) and (D), sub-Poissonian behavior exists for all the range of $J>0$, and $\alpha,\beta>-1$.
  \item The reader will observe that the frames A(d) and B(b), as well as A(a) and B(d) are very similar, respectively, but not identical despite different values for $\alpha$ and $\beta$. However one has no rigorous answer to this remark.
  \item We noticed that the transition quoted at the first point (i) occurs if and only if the restriction $|\alpha+\beta|=1$ holds, as shown in the frames A(b) and B(c).
  \item Finally, we observed in all frames that the behavior of the Mandel parameter reaches $-\frac{1}{2}$ as $J\rightarrow\infty$.
\end{enumerate}

\indent The last observation can be mathematically explained using the asymptotic forms, as $J\rightarrow\infty$, of the modified Bessel
functions $I_\nu(z)$ (see the identity 14.143, pp. 693 of \cite{57})
\begin{eqnarray}\label{4.11}
  I_\nu(z) &=& \frac{\rme^z}{\sqrt{2\pi z}}\left(\mathbf P_\nu(\rmi z)-\rmi\,\mathbf Q_\nu(\rmi z)\right),
\end{eqnarray}
valid for $-\pi/2<{\rm arg} z<\pi/2$, where $\mathbf P_\nu(\rmi z)$ and $\mathbf Q_\nu(\rmi z)$ are defined through
\begin{eqnarray*}
\fl  \mathbf P_\nu(\rmi z) \sim 1-\frac{(4\nu^2-1)(4\nu^2-9)}{2!(8z)^2}
                               +\frac{(4\nu^2-1)(4\nu^2-9)(4\nu^2-25)(4\nu^2-49)}{4!(8z)^4}-\cdots, \nonumber\\
\fl  \mathbf Q_\nu(\rmi z) \sim \frac{4\nu^2-1}{1!(8z)}-\frac{(4\nu^2-1)(4\nu^2-9)(4\nu^2-25)}{3!(8z)^3}+\cdots. \nonumber
\end{eqnarray*}
\indent For larger $z\equiv2\sqrt{2J}$, i.e. as $J\rightarrow\infty$, the first terms in $\mathbf P_\nu(\rmi z)$ and $\mathbf Q_\nu(\rmi z)$ dominate, and thus it is convenient to rewrite \eref{4.11} as
\begin{eqnarray*}
  I_\nu(2\sqrt{2J}) &\sim&  \frac{\rme^{2\sqrt{2J}}}{2\sqrt{\pi\sqrt{2J}}}\left(1-\frac{4\nu^2-1}{16\sqrt{2J}}\right),
\end{eqnarray*}
and substituting $\nu=(2\sigma,2\sigma+1,2\sigma+2)$ in the last identity, \eref{4.9} becomes
\begin{eqnarray*}
  Q_{\rm M}^{(m)}(J) &\sim& -\frac{1}{2}-\frac{(4\sigma+1)(4\sigma+3)}{64}\sqrt{\frac{2}{J}},\qquad(\sigma=s-m),
\end{eqnarray*}
where in the limit $J\rightarrow\infty$, the second term in the last expression can be neglected and one is left with the Mandel parameter which tends to $-\frac{1}{2}$, for $\forall\sigma\in\mathbb R$, as illustrated in \Fref{figure3}.\\
\indent As a prerequisite for obtaining quantum revivals, it has been shown that a coherent state wave-packet of the form of \eref{4.6} mimics quantum revivals, $T^{(m)}_{\rm rev}=4\pi/|e''^{(m)}_{\overline n}|$, and fractional revivals $\tau=(p/q)T_{\rm rev}$, in which $p$ and $q$ are coprime integers, besides the classical timescale $T^{(m)}_{\rm cl}=2\pi/|e'^{(m)}_{\overline n}|$ if and only if they are strongly well localized around a mean value $\overline n\simeq\langle n\rangle$. This means that we can expand the energy eigenvalues \eref{2.16} in Taylor series in $n$ as
\begin{eqnarray}\label{4.12}
  e^{(m)}_n=\frac{1}{2}\left[\overline n(\overline n+2\sigma)+\frac{4\pi}{T^{(m)}_{\rm cl}}(n-\overline n)
                             +\frac{2\pi}{(\overline n+\sigma)T^{(m)}_{\rm cl}}(n-\overline n)^2\right],
\end{eqnarray}
with $k=1$ and the timescales are given by $T^{(m)}_{\rm cl}=2\pi/(\overline n+\sigma)$ and $T^{(m)}_{\rm rev}=4\pi,\,\forall m\geq0$.\\
\indent Taking into account \eref{4.12}, the Gazeau-Klauder CS \eref{4.5} reads as
\begin{eqnarray}\label{4.13}
\fl  |\xi^{(m)}(x;J,\overline t)\rangle=&\frac{(2J)^{\sigma/2}}{\sqrt{I_{2\sigma}(2\sqrt{2J})}}\,
\rme^{-\rmi\,\omega\,\overline n\left(\pi+\frac{\sigma}{2}T^{(m)}_{\rm cl}\right)\overline t} \nonumber \\
&    \times\sum_{n=0}^\infty\frac{(2J)^{n/2}\exp\left\{-\rmi\,\omega\,\pi(n-\overline n)\left(2+\frac{n-\overline n}{\overline n+\sigma}\right)\,\overline t\right\}}{\sqrt{n!\,\Gamma(n+2\sigma+1)}}\,|\psi^{(m)}_n(x)\rangle,
\end{eqnarray}
where $\overline t=t/T^{(m)}_{\rm cl}$.\\
\indent Generally, the autocorrelation function, $A^{(m)}(t)=\langle\xi^{(m)}(x;J,0)|\xi^{(m)}(x;J,t)\rangle$, and the probability density of the time-evolved coherent state wave-packet are considered as widely used techniques for describing and reproducing revival structures. To this end using the product series relation {\bf 0.316} of \cite{58}, the absolute square of $A^{(m)}(\overline t)$ and the probability
density of \eref{4.13} yield
\begin{eqnarray}
\fl  |A^{(m)}(\overline t)|^2=\left(\frac{(2J)^{\sigma}}{I_{2\sigma}(2\sqrt{2J})}\right)^2\sum_{n=0}^\infty\sum_{l=0}^n\frac{(2J)^{n}
\exp\left\{-\rmi\,\omega\,\pi\frac{(n-2l)(n+2\sigma)}{\overline n+\sigma}\,\overline t\right\}}{l!\,(n-l)!\,\Gamma(l+2\sigma+1)\,
\Gamma(n-l+2\sigma+1)},\label{4.14} \\
\fl  |\xi^{(m)}(x;J,\overline t)|^2=\frac{(2J)^{\sigma}}{I_{2\sigma}(2\sqrt{2J})} \nonumber \\
 \fl \qquad\quad\times\sum_{n=0}^\infty
\sum_{l=0}^n\frac{(2J)^{n/2}\exp\left\{-\rmi\,\omega\,\pi\frac{(n-2l)(n+2\sigma)}{\overline n+\sigma}\,\overline t\right\}
}{\sqrt{l!\,(n-l)!\,\Gamma(l+2\sigma+1)\,\Gamma(n-l+2\sigma+1)}}\,\left(\psi^{(m)}_l(x)\right)^\ast\psi^{(m)}_{n-l}(x),\label{4.15}
\end{eqnarray}
where the eigenfunctions $\psi^{(m)}_n(x)$ are given in \eref{2.17}.\\
\indent Both expressions \eref{4.14} and \eref{4.15} are attributed to the exceptional $X_m$ PDEM Scarf I potential and we use them to study their revival dynamics. However, it is obvious that all these results will reduce to those of the usual constant mass (CM) if we set $M(x)=1$, i.e., $\mu(x)=x$. At this stage our strategy in the remainder of this paper is as follows. To illustrate how \eref{4.14} and \eref{4.15} work, we start by studying the usual CM in which revival structures are well known and this in order to confirm the validity of our assertions. Once the results are verified, we then apply both expressions to the exceptional $X_m$ PDEM system, with an appropriate choice of the mass function $M(x)$, to see what general conclusions can be made.\\
\indent Throughout our results, we work in atomic units (a.u.), i.e., $\hbar=e=m_0=\omega=1$, for which
the conversions give $1\,{\rm a.u.}\, \simeq5.29\times10^{-10}$ m for lengths and $1\,{\rm a.u.}\,\simeq2.42\times10^{-17}$ sec for times. For convenience we also set $k=1$.

\subsection{The case of constant mass}%

\noindent According to the discussion above, coherent state wave-packet of the Scarf I potential have perfect full revivals since their
energy spectrum is quadratic in the quantum number $n$. We illustrate this by plotting the absolute square of $A^{(m)}(\overline t)$ as a
function of $\overline t=t/T^{(m)}_{\rm cl}$ and the modulus square of $\xi^{(m)}(x;J,\overline t)$ against $x$.

\begin{figure}[h]
  \includegraphics[width=15.5cm,height=12.5cm]{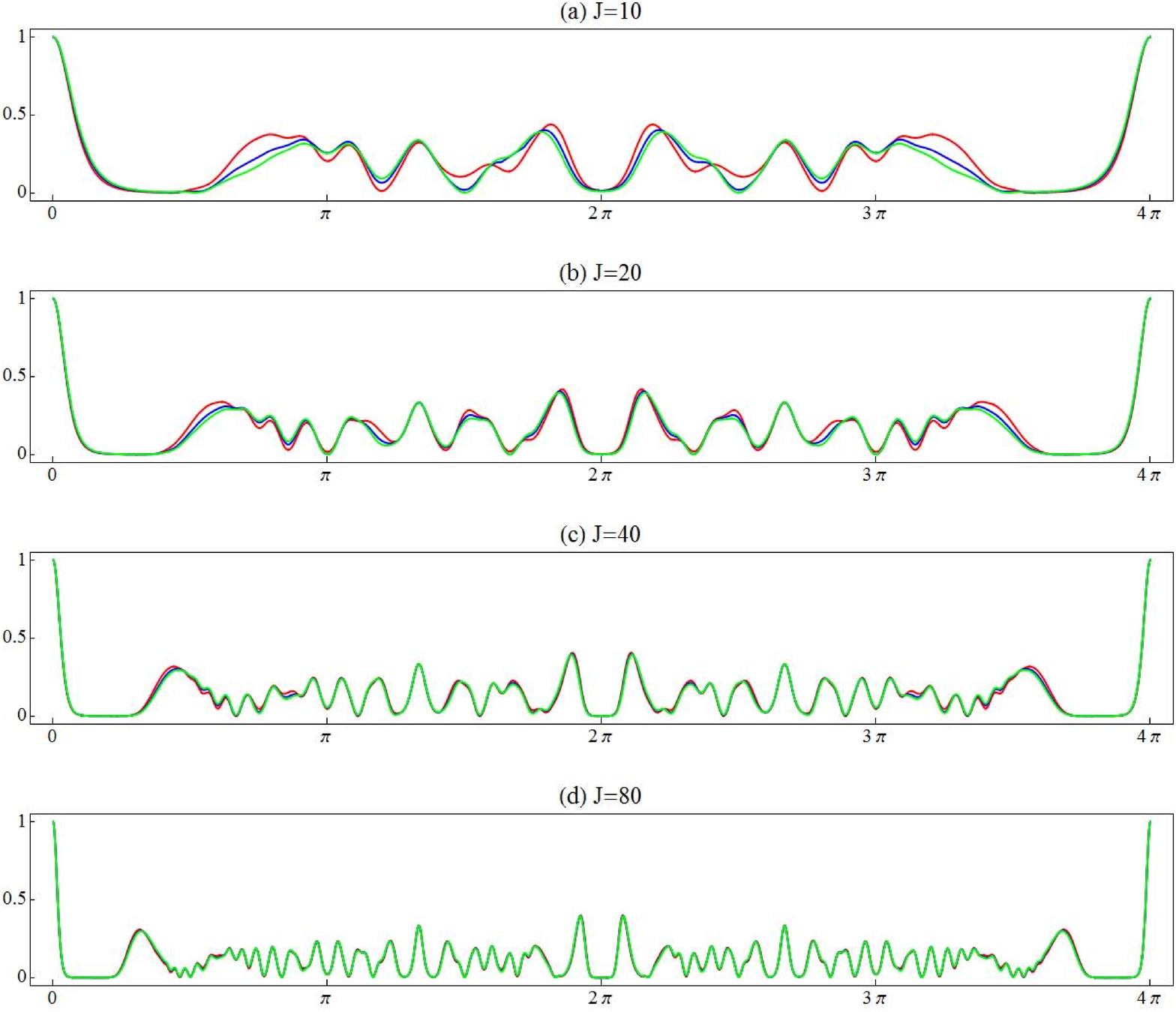}
  \caption{The modulus square of autocorrelation function $A^{(m)}(\overline t)$ against $\overline t=t/T^{(m)}_{\rm cl}$ plotted for
  $\alpha=3/2$, $\beta=5/2$, $n_{\rm max}=50$. Here the parameter $m$ refers to: $m=0$ (red line), $m=1$ (blue line) and $m=2$ (green line).}
  \label{figure4}
\end{figure}

\indent \Fref{figure4} shows the evolution of $|A^{(m)}(\overline t)|^2$ for $J=10$, $20$, $40$, and $80$, with $\alpha=\frac{3}{2},
\beta=\frac{5}{2},n_{\rm max}=50$ (i.e., 50th excited state) for the classical ($m=0$) and exceptional ($m=1,2$) PDEM Scarf I potential. The timescales are $T^{(m)}_{\rm rev}=4\pi$ and $T^{(m)}_{\rm cl}=2\pi/(\overline n+\sigma)$. As we can see, the usual CM involves a perfect full revivals as it was expected above and the sharp peaks arise due to the fractional revivals which become more apparent as $J$ increases. However some permanent peaks still exist, no matter the values attributed to $J$ and $m$, like those of $T_{\rm rev}/3$ and $2T_{\rm rev}/3$, for all $m=0,1,2$, with a change in the width and magnitude. We also observe that as $J$ increases all curves merge into a single one, as can be seen in \Fref{figure4}(d).

\begin{figure}[h]
  \includegraphics[width=16.3cm,height=3cm]{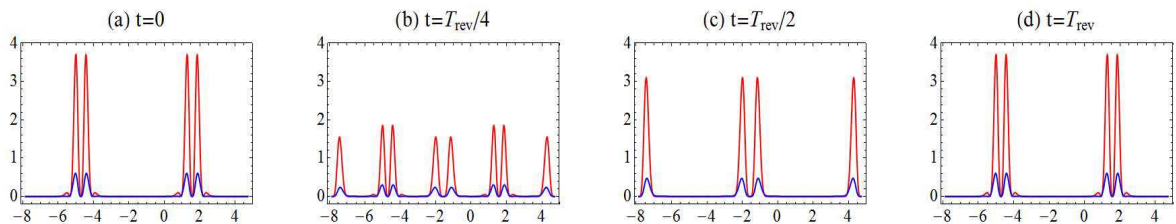}
  \caption{Plot of $|\xi^{(m)}(x;J,\overline t)|^2$ against $x$ for $J=20$, plotted for $\alpha=1$, $\beta=2$, and $n_{\rm max}=50$. As for \Fref{figure4}, the parameter $m$ refers to: $m=0$ (red line) and $m=1$ (blue line).}\label{figure5}
\end{figure}

\indent On the other hand we plot in \Fref{figure5} the probability density of a coherent state wave-packet $|\xi^{(m)}(x;J,\overline t)|^2$ for $J=20$, $\alpha=1,\,\beta=2$ and $m=0,1$. It is evident that the probability density is reconstructed after the time revival
$T^{(m)}_{\rm rev}$ as it was illustrated in the frame (d) compared to (a). In the first frame, $t=0$, we observe two principal and
symmetrical peaks for both cases, ($m=0,1$), centered at $x_1\simeq1.25$ and $x_2\simeq2$, followed by a secondary ripple in the case $m=0$ before and after dominant peaks. All these phenomena are periodic in the whole real line with a period $\delta x=2\pi$, being the behavior for the case where the mass is constant. With time, our numerical simulations show us that these peaks oscillate back and forth between the walls of the well with relative change in shape.

\subsection{The case of position-dependent effective mass}%

\noindent Even as \eref{4.14} and \eref{4.15} are applicable for the exceptional $X_m$ PDEM Scarf I potential, two different profiles of the mass function, introduced in section 2, without and with singularities were chosen
\begin{eqnarray}\label{4.16}
  M_{\rm wos}(x)=\frac{1}{1+(\lambda x)^2},\qquad{\rm and} \qquad M_{\rm ws}(x)=\frac{1}{\left(1-(\lambda x)^2\right)^2},
\end{eqnarray}
respectively. These mass functions allow us to construct a coherent state wave-packet with the proper behavior near the boundaries and have been used in many studies, see for instance \cite{59}.

\subsubsection{Mass function $M_{\rm wos}(x)$ without singularities.}

\begin{figure}[h]
  \includegraphics[width=16.3cm,height=9cm]{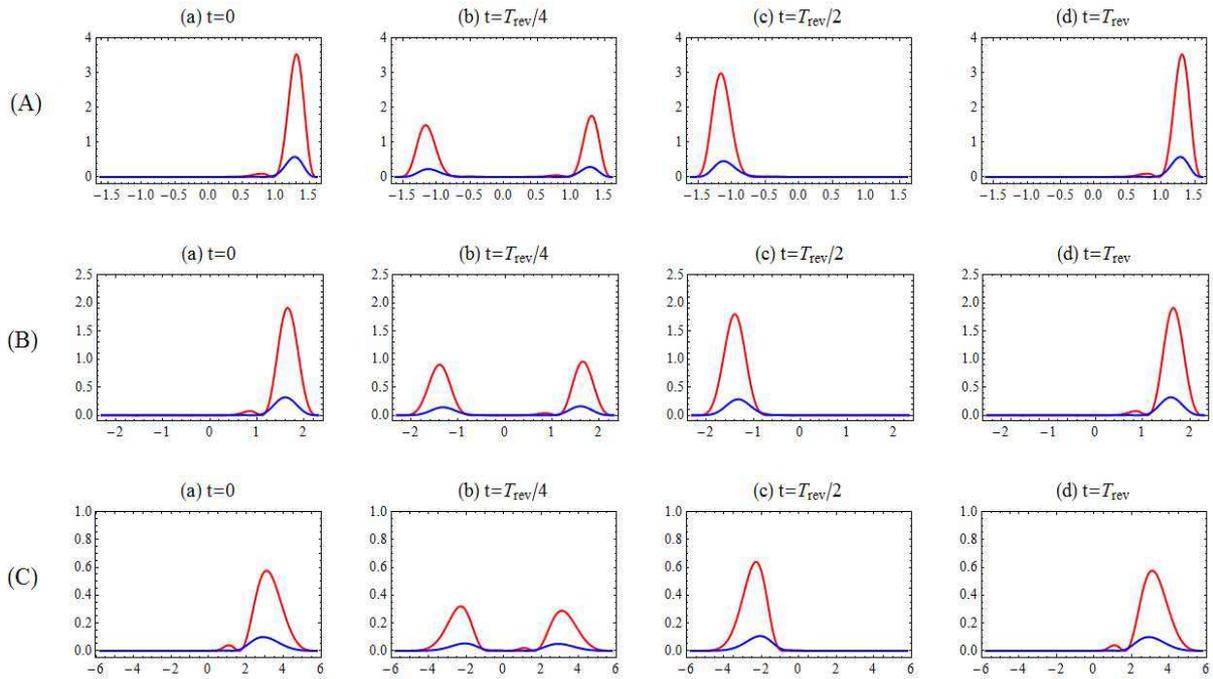}
  \caption{Plot of $|\xi^{(m)}(x;J,\overline t)|^2$ against $x$ for the profile of the mass function $M_{\rm wos}(x)$, with $\alpha=1$, $\beta=2$, $J=20$, $n_{\rm max}=50$. The mass parameters $\lambda$ are: (A) $\lambda=0.25$, (B) $\lambda=1$, and (C) $\lambda=2$, and the parameter $m$ refers to: $m=0$ (red line) and $m=1$ (blue line).}\label{figure6}
\end{figure}

We take the profile of the mass function to be of the form of $M_{\rm wos}(x)$. This profile is without singularities and is a bounded function defined in the whole real line $\mathbb R$, where its maximum value, $M_{\rm wos}^{(\rm max)}(x)=1$, is reached at $x=0$ and vanishing as $|x|\rightarrow\infty$. In this case the auxiliary mass function is calculated by a simple integration, which gives
$\mu_{\rm wos}(x)=\frac{1}{\lambda}\,\arcsh(\lambda x)$.\\
\indent In \Fref{figure6}, we plot the probability density of a coherent state wave-packet endowed with an effective mass function $M_{\rm wos}(x)$ for different values of the mass parameter $\lambda$, taking into account \eref{2.18}, i.e.,
$|x|\leq\frac{1}{\lambda}\sinh\left(\frac{\pi\lambda}{2}\right)$. The revival time is $T^{(m)}_{\rm rev}=4\pi$ and we observe that all
$|\xi^{(m)}(x;J,\overline t)|^2$ are restored after the time revival as they are presented in the frames A-C(d). With the presence of the
mass function, we can see that the dependence of the mass on the position $x$ affects wholly the behavior of coherent state wave-packets
inside the corresponding wells through two ways: firstly we observe that, although full revivals take place during time evolution
$T^{(m)}_{\rm rev}$, there is no trace of fractional revivals in the common sense on the opposition of the usual CM discussed above.
Secondly the symmetrical peaks of the \Fref{figure5} do not occur in the case of mass $M_{\rm wos}(x)$, instead we observe only one peak, for each cases $m=0,1$, with the same width than the usual which is nearly zero at the left of the well and the formation of ripple before it. As time evolves we observe a well localized coherent state wave-packet oscillating back and forth between the walls, with the presence of two asymmetrical peaks at $T^{(m)}_{\rm rev}/4$ known as mirror revivals, as we can see in the frames A-C(b). Gradually as the mass parameter $\lambda$ increases, a coherent state wave-packet spreads on the whole real line inside the region delimited by the walls of the potential well. This behavior is due essentially to the features of the mass function.

\begin{figure}[h]
  \includegraphics[width=16.3cm,height=9cm]{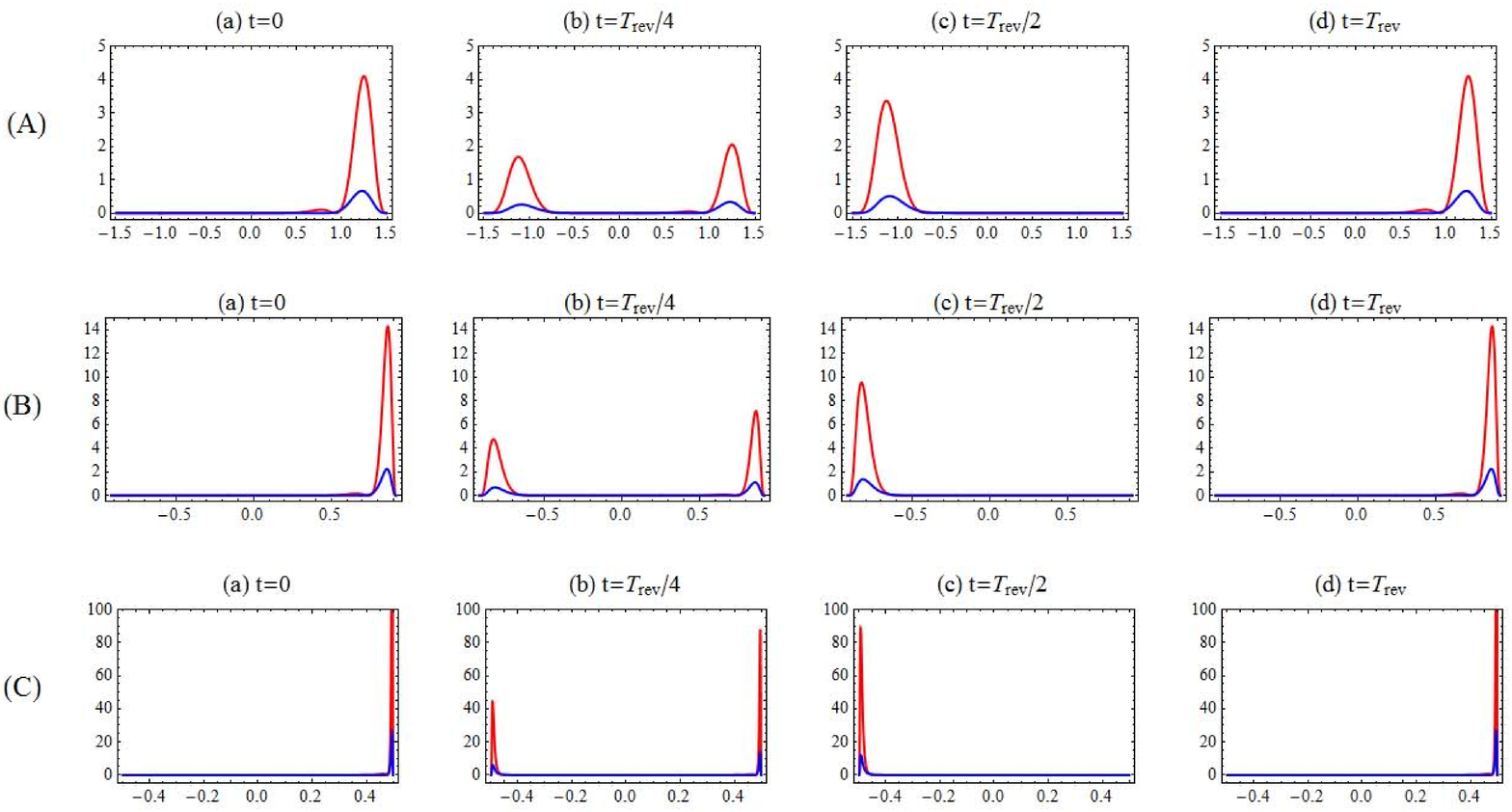}
  \caption{Plot of $|\xi^{(m)}(x;J,\overline t)|^2$ against $x$ for the profile of the mass function $M_{\rm ws}(x)$, with $\alpha=1$, $\beta=2$, $J=20$, $n_{\rm max}=50$. The mass parameters $\lambda$ are: (A) $\lambda=0.25$, (B) $\lambda=1$, and (C) $\lambda=2$, and $m$ refers to: $m=0$ (red line) and $m=1$ (blue line).}\label{figure7}
\end{figure}

\subsubsection{Mass function $M_{\rm ws}(x)$ with two singularities.}
The second mass function is chosen to be of the form of $M_{\rm ws}(x)$ with two singularities and defined in
${\rm dom}(M_{\rm ws})=(-1/\lambda,+1/\lambda)$. The mass function rapidly grows near the classical turning points $x_\pm=\pm1/\lambda$ and reaches its minimum at $x_0=0$. The associated auxiliary mass function is given by $\mu_{\rm ws}(x)=\frac{1}{\lambda}\,\arcth(\lambda x)$.\\
\indent In \Fref{figure7} we display the probability density of a coherent state wave-packet with the mass function $M_{\rm ws}(x)$ for different values of $\lambda$, with the restriction \eref{2.18} given by $|x|\leq\frac{1}{\lambda}\tanh\left(\frac{\pi\lambda}{2}\right)$. With this profile at hand, an analogous temporal evolution takes place in the \Fref{figure7}(A) compared to those of \Fref{figure6}(A). This is essentially due to the fact that both hyperbolic functions "$\sinh$" and "$\tanh$" behave in the same manner as $\lambda$ approaches zero and one can say that the same quantitative comments can be made in the first case with an exception that a new phenomenon is observed here. Contrary to the case of the mass $M_{\rm wos}(x)$, a coherent state wave-packet in frames (B) and (C) becomes more peaked and tends to gather near the classical turning points $x_\pm=\pm1/\lambda$ of the well due to singularities of the mass function and progressively becomes more sharper than the usual ones as $\lambda$ increases. We see also that the amplitude of the probability density of a coherent state wave-packet rapidly grows as one approaches the classical turning points. It is clear that the presence of singularities in the mass function restrict the time evolution of a coherent state wave-packet inside the domain determined by the turning points $x_\pm$, once again this is due to the features of the mass function.

\begin{figure}[h]
  \includegraphics[width=16cm,height=6cm]{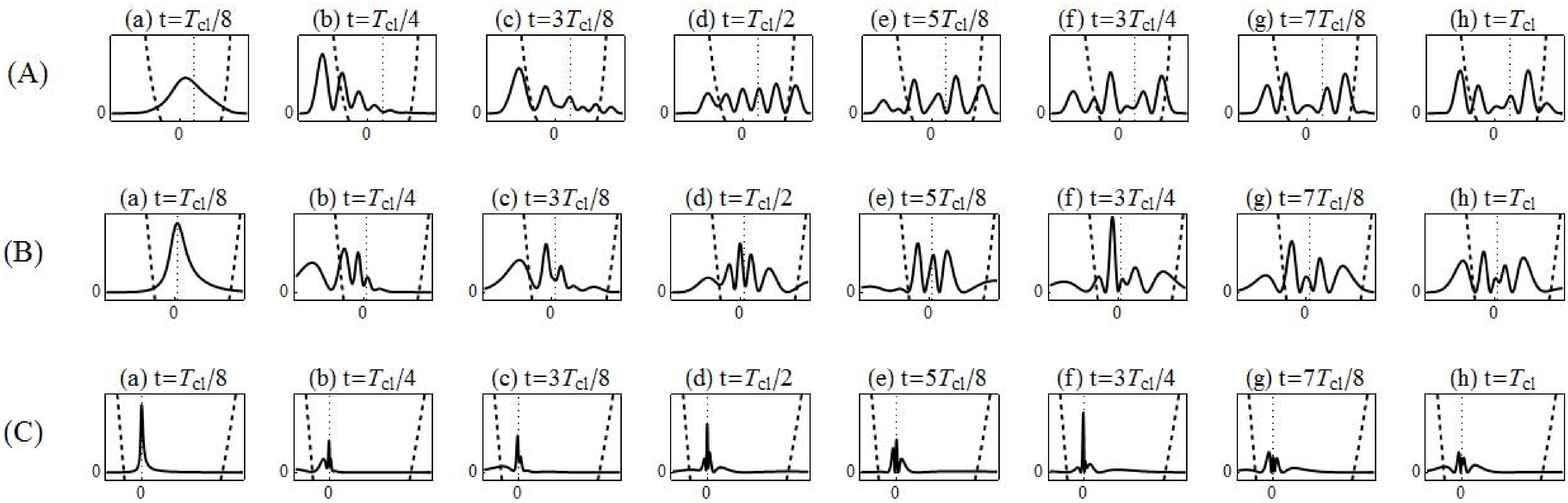}
  \caption{\footnotesize{Time evolution of the probability density of a coherent state wave-packet for the mass function $M_{\rm wos}(x)$ for $m=0$, with $\alpha=1$, $\beta=2$, $J=20$, and $n_{\rm max}=50$. The mass parameters $\lambda$ are: (A) $\lambda=0.5$, (B) $\lambda=2$, and (C) $\lambda=4$.}}\label{figure8}
\end{figure}
\begin{figure}[h]
  \includegraphics[width=16cm,height=6cm]{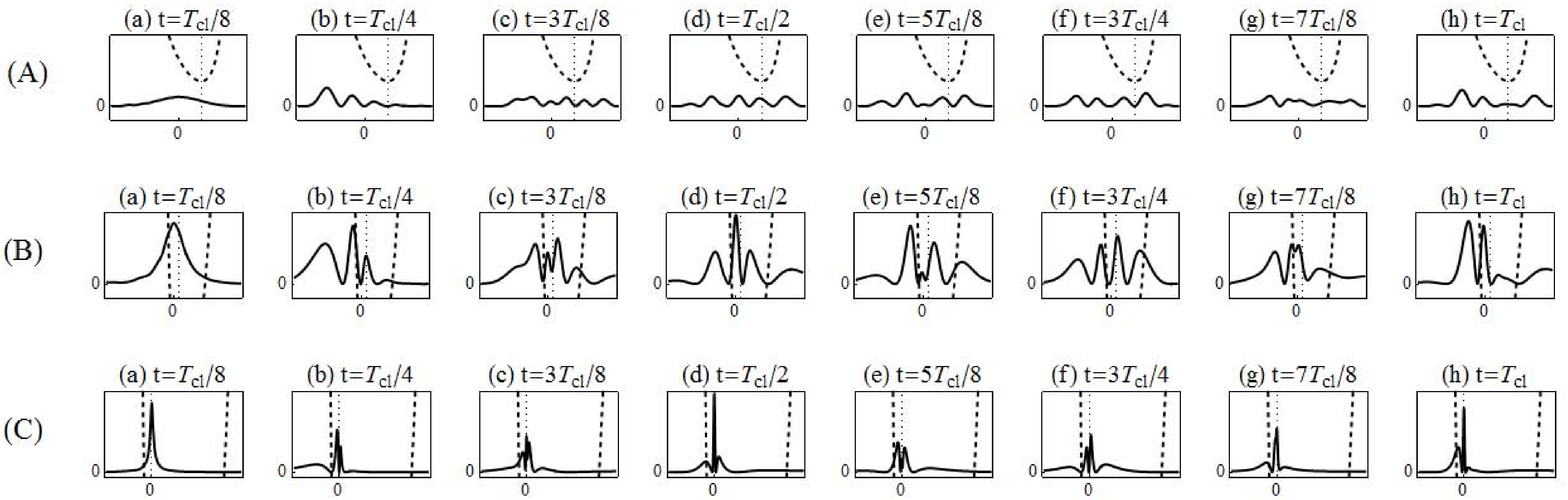}
  \caption{\footnotesize{Time evolution of the probability density of a coherent state wave-packet for the mass function $M_{\rm wos}(x)$ for $m=1$, with $\alpha=1$, $\beta=2$, $J=20$, and $n_{\rm max}=50$. The mass parameters $\lambda$ are: (A) $\lambda=0.5$, (B) $\lambda=2$, and (C) $\lambda=4$.}}\label{figure9}
\end{figure}

\indent We end our analysis by showing the time evolution of the probability density of a coherent state wave-packet
$|\xi^{(m)}(x;J,\overline t)|^2$ for the exceptional $X_m$ PDEM Scarf I potential, characterized by not-equally spaced eigenenergies
\eref{2.16}. It is well-known, as it was exposed in the pedagogical paper of Gutschick and Nieto \cite{54}, that for a system with a such
eigenenergies, coherent state wave-packets will dissipate and lose their coherence in time. Thus the concept of coherence is discussed in our paper in terms of classical period $T^{(m)}_{\rm cl}=2\pi/(\overline n^{(m)}+\sigma)$ and defined as follow \cite{54}: {\it the more eigenstates $\psi_n^{(m)}(x)$ have a significant overlap with the PDEM Gazeau-Klauder CS in between the walls of the potential $V_{\rm eff}^{(m)}(x)$, the longer will be the coherence time $\tau^{(m)}_{\rm coh}$}. To this end, \Fref{figure8} and \Fref{figure9} display the time evolution of probability densities of a PDEM coherent state wave-packet \eref{4.15} for $m=0$ and $m=1$, respectively, for the profile mass function $M_{\rm wos}(x)$, given in \eref{4.16}, over one classical period $T^{(m)}_{\rm cl}$ for $\lambda=0.5,2$, and $4$ in the case $J=20$. All frames are taken as $\frac{1}{8}$th of one classical period and show: (i) $|\xi^{(m)}(x;J,\overline t)|^2$ (solid curves), (ii) the associated potential (dashed curves), and (iii) a vertical dotted line indicating the potential minimum.
\begin{description}
  \item[Case $m=0$.] Numerical simulations show us that the exceptional $X_0$ PDEM coherent state wave-packet, associated to the classical Jacobi polynomials, starts their movement to the right of the potential minimum and oscillates back and forth inside the well. We observe in \Fref{figure8}(A) a shorter coherence time in terms of classical period which means that, for $\lambda=0.5$, the system loses its coherence quickly as it was represented in the frame A(b). In \Fref{figure8}(B), gradually as $\lambda$ increases, $\lambda=2$, a coherent state wave-packet continues to lose its coherence but slowly compared to the previous case. In both cases, we see that the coherence time is {\it less} than one classical period, i.e., $\tau^{(0)}_{\rm coh}<T^{(0)}_{\rm cl}\simeq 0.98118$, for $\langle n^{(0)}\rangle\sim\overline n^{(0)}\simeq4.40365$.\\
      However in \Fref{figure8}(C), corresponding to $\lambda=4$, we observe that PDEM coherent state wave-packets are more peaked at the vicinity of the potential minimum and flatten out very quickly to reach zero at the right of the potential well. The eigenstates and PDEM CS wave-packets overlap very well, and these phenomena are the signature of a longer coherence time, {\it at least longer} than one classical period, i.e., $\tau^{(0)}_{\rm coh}>T^{(0)}_{\rm cl}$.
  \item[Case $m=1$.] In \Fref{figure9}(A), corresponding to $\lambda=0.5$, the remark which should be emphasized is that we observe a
      {\it strong loss} of coherence since, due to the definition herein above, eigenstates which are outside of the well do not overlap with the PDEM CS wave-packets. However, the slow loss of coherence in \Fref{figure9}(B) is very similar to that of \Fref{figure8}(B), with $\tau^{(1)}_{\rm coh}<T^{(1)}_{\rm cl}\simeq 1.02087$ for $\langle n^{(1)}\rangle\sim\overline n^{(1)}\simeq5.15475$.\\
      Finally, all the qualitative comments made for \Fref{figure8}(C) also hold for \Fref{figure9}(C), since the reader can easily observe that both frames exhibit the same phenomenon, i.e., longer coherence time.
\end{description}

\indent Thus, we suspect the effect that the loss of coherence depends closely on the mass parameter $\lambda$ and we finish before our
conclusion with this observation: {\it a larger mass parameter $\lambda$ contributes significantly to a longer coherence time}
$\tau^{(m)}_{\rm coh}$.

\section{Conclusion}%

\noindent In this paper our primary concern is to investigate how the mass function, represented here by the parameter $\lambda$, can affect the revival structure of an arbitrary quantum system. To this end we have constructed the PDEM Gazeau-Klauder coherent states for the exceptional $X_m$ Scarf I potential endowed with PDEM, where their statistical and dynamical properties have been studied. We have shown that these potentials are shape invariant and are isospectral to PDEM potentials whose solutions are given in terms of the classical Jacobi polynomials ($m=0$). In particular, for the usual CM, we have constructed full and fractional revivals with the help of the autocorrelation function. However, in the case of PDEM, things are completely different from the usual case and our results agree with those obtained by Schmidt in \cite{51}. We have observed that, although full revivals still take place during their time evolution $T^{(m)}_{\rm rev}$, there is no trace of fractional revivals in the common sense, on the opposition to the usual. Instead of these effects, we have obtained bell-shaped coherent state wave-packets located in the right of the well, oscillating back and forth between the walls. We have concluded that not only quantum revivals are different and affected but also depend closely on the profile of the mass function. In this context two profiles were chosen, with and without singularities, to illustrate numerically the dynamic of their revival structures.\\
\indent We have also observed that for a longer coherence time $\tau^{(m)}_{\rm coh}$, defined here in the sense of a slow loss of coherence, corresponds a larger mass parameter $\lambda$. Then, we suspect the effect that $\lambda$ affects considerably $\tau^{(m)}_{\rm coh}$, which leads the state to lose its coherence in time more (resp. less) rapidly as $\lambda$ becomes smaller (resp. larger).

\section*{References}%


\begin{thebibliography}{99}

\bibitem{1} Harrison P 2005 {\it Quantum Wiles, Wires and Dots. Theoretical and Computational Physics of Semiconductior Nanostructures}
(New York: John Wiley $\&$ Sons, LTD)
\bibitem{2} Bastard G 1998 {\it Wave Mechanics Applied to Semiconductor Heterostructures} (France: les \'Editions de Physique, les Ullis)
\bibitem{3} Ring P and Schuck P 1980 {\it The Nuclear Many Boby Problems} (New York: Springer)
\bibitem{4} Quesne C and Tkachuk V M 2004 {\it J. Phys. A: Math. Gen.} {\bf 37} 4267
\bibitem{5} Jiang L, Yi L-Z and Jia C-S 2005 {\it Phys. Lett. A} {\bf 345} 279
\bibitem{6} Mustafa O and Mazharimousavi S H 2008 {\it J. Phys. A: Math. Theor.} {\bf 41} 244020 (and references therein.)
\bibitem{7} Ruby V C and Senthilvelan M 2010 {\it J. Math. Phys.} {\bf 51} 52106
\bibitem{8} Cruz y Cruz S and Rosas-Ortiz O 2009 {\it J. Phys. A: Math. Theor.} {\bf 42} 185205
\bibitem{9} Amir N and Iqbal S 2015 {\it J. Math. Phys.} {\bf 56} 062108
\bibitem{10} Yahiaoui S-A and Bentaiba M 2014 {\it J. Phys. A: Math. Theor.} {\bf 47} 025301
\bibitem{11} Yahiaoui S-A and Bentaiba M 2012 {\it J. Phys. A: Math. Theor.} {\bf 45} 444034
\bibitem{12} Cherroud O, Yahiaoui S-A and Bentaiba M 2016 {\it J. Math. Phys.} (submitted)
\bibitem{13} Chen Z D and Chen G 2006 {\it Phys. Scr.} {\bf 73} 354
\bibitem{14} de Souza Dutra A and de Oliveira J A 2008 {\it Phys. Scr.} {\bf 78} 035009
\bibitem{15} Dong S H 2007 {\it Factorization Methods in Quantum Mechanics} (Dordrecht: Springer)
\bibitem{16} Cooper F, Khare A and Sukhtame U 2001 {\it Supersymmetry and Quantum Mechanics} (Singapore: World scientific)
\bibitem{17} Iachello F 2006 {\it Lie Algebra and Applications}, Lect. Notes Phys. 708 (Berlin: Springer)
\bibitem{18} Yahiaoui S-A and Bentaiba M 2009 {\it Int. J. Theor. Phys.} {\bf 48} 315
\bibitem{19} Ko\c{c} R and Koca M 2003 {\it J. Phys. A: Math. Theor.} {\bf 36} 8105
\bibitem{20} Ju G-X, Cai C-Y, Xiang Y and Ren Z-Z 2007 {\it Commun. Theor. Phys.} {\bf 47} 1001
\bibitem{21} Ismail M E H and van Assche W 2005 {\it Classical and Quantum Orthogonal Polynomials in One Variable} (Cambridge: Cambridge
Univ. Press)
\bibitem{22} G\'omez-Ullate D, Kamran N and Milson R 2009 {\it J. Math. Anal. Appl.} {\bf 359} 352
\bibitem{23} G\'omez-Ullate D, Kamran N and Milson R 2010 {\it J. Approx. Theor.} {\bf 162} 987
\bibitem{24} Quesne C 2008 {\it J. Phys. A: Math. Theor.} {\bf 41} 392001
\bibitem{25} Fellows J M and Smith R A 2009 {\it J. Phys. A: Math. Theor.} {\bf 42} 335303
\bibitem{26} Quesne C 2009 {\it SIGMA} {\bf 5} 084
\bibitem{27} Ho C-L 2011 {\it J. Math. Phys.} {\bf 52} 122107
\bibitem{28} Ho C-L, Odake S and Sasaki R 2011 {\it SIGMA} {\bf 7} 107
\bibitem{29} G\'omez-Ullate D, Kamran N and Milson R 2012 {\it Contemp. Math.} {\bf 563} 51
\bibitem{30} Quesne C 2012 {\it SIGMA} {\bf 8} 080
\bibitem{31} G\'omez-Ullate D, Kamran N and Milson R 2013 {\it Found. Comput. Math.} {\bf 13} 615
\bibitem{32} Marquette I and Quesne C 2013 {\it J. Phys. A: Math. Theor.} {\bf 46} 155201
\bibitem{33} Odake S and Sasaki R 2013 {\it J. Phys. A: Math. Theor.} {\bf 46} 245201
\bibitem{34} Odake S and Sasaki R 2013 {\it J. Phys. A: Math. Theor.} {\bf 46} 235205
\bibitem{35} Grandati Y and Quesne C 2013 {\it J. Math. Phys.} {\bf 54} 073512
\bibitem{36} G\'omez-Ullate D, Grandati Y and Milson R 2014 {\it J. Phys. A: Math. Theor.} {\bf 47} 015203
\bibitem{37} G\'omez-Ullate D, Grandati Y and Milson R 2014 {\it J. Math. Phys.} {\bf 55} 043510
\bibitem{38} Quesne C 2015 {\it J. Phys.: Conf. Ser.} {\bf 597} 012064
\bibitem{39} Midya B and Roy B 2009 {\it Phys. Lett. A} {\bf 373} 4117
\bibitem{40} Robinett R W 2004 {\it Phys. Rep.} {\bf 392} 1
\bibitem{41} For primers on full and fractional revivals, see Bluhm R, Kostelech\'y V A and Porter J A 1996 {\it Am. J. Phys.} {\bf 64} 944
\bibitem{42} Romera E and de los Santos F 2009 {\it Phys. Rev. B} {\bf 80} 165416
\bibitem{43} Torres J J and Romera E 2010 {\it Phys. Rev. B} {\bf 82} 155419
\bibitem{44} Demikhovskii V Ya, Maksimova G M, Perov A A and Telezhnikov A V 2012 {\it Phys. Rev. A} {\bf 85} 022105
\bibitem{45} Krueckel V and Kramer T 2009 {\it New J. Phys.} {\bf 11} 093010
\bibitem{46} Schliemann J 2008 {\it New J. Phys.} {\bf 10} 043024
\bibitem{47} Garcia T, Rodr\'iguez-Bolivar S, Cordera N A and Romera E 2013 {\it J. Phys.: Condens. Matter} {\bf 25} 235301
\bibitem{48} de los Santos F and Romera E 2013 {\it Phys. Rev. A} {\bf 87} 013424
\bibitem{49} Romera E and de los Santos F 2013 {\it Phys. Lett. A} {\bf 377} 2284
\bibitem{50} Dey S, Fring A, Gouba L and Castro P G 2013 {\it Phys. Rev. D} {\bf 87} 084033
\bibitem{51} Schmidt A G M 2006 {\it Phys. Lett. A} {\bf 353} 459
\bibitem{52} Midya B, Roy B and Biswaz A 2009 {\it Phys. Scr.} {\bf 79} 065003
\bibitem{53} Gazeau J-P 2009 {\it Coherent states in Quantum Mechanics} (Berlin: Wiley-VCH)
\bibitem{54} Gutschick V P and Nieto M M 1980 {\it Phys. Rev. D} {\bf 22} 403
\bibitem{55} Midya B and Roy B 2013 {\it J. Phys. A: Math. Theor.} {\bf 46} 175201
\bibitem{56} BenDaniel D J and Duke C B 1966 {\it Phys. Rev. B} {\bf 152} 683
\bibitem{57} Arfken G B, Weber H J and Harris F E 2013 {\it Mathematical Methods for Physicists. A Comprehensive Guide}, 7th. edition
(New York: Academic Press)
\bibitem{58} Gradshteyn I S and Ryzhik I M 2007 {\it Table of Integrals, Series and Products} (New York: Academic Press)
\bibitem{59} Cruz y Cruz S, Negro J and Nieto L M 2007 {\it Phys. Lett. A} {\bf 369} 400
\bibitem{60} Bagchi B, Banerjee B, Quesne C and Tkachuk V M 2005 {\it J. Phys. A: Math. Gen.} {\bf 38} 2929

\end{thebibliography}
\end{document}